\documentclass[aps,twocolumn,showpacs,preprintnumbers, floats]{revtex4}
\usepackage{amsfonts}
\usepackage{amsmath}
\usepackage{graphicx}
\usepackage{dcolumn}
\usepackage{bm}

\begin{document}

\preprint{}
\title{Tuning the atomic and domain structure of epitaxial films of
  multiferroic BiFeO$_3$}
\author{C.J.M. Daumont,$^{1}$ S. Farokhipoor,$^{1}$ A. Ferri,$^{1}$
  J. C. Wojde\l,$^{2}$ Jorge I\~{n}iguez,$^{2}$ B.J. Kooi$^{1}$}
\author{B. Noheda$^{1}$}
\email{b.noheda@rug.nl, Corresponding author}
\affiliation{$^{1}$Zernike Institute for Advanced Materials, University of Groningen, Groningen 9747AG, The Netherlands}
\affiliation{$^{2}$Institut de Ci\`{e}ncia de Materials de Barcelona (CSIC), Campus UAB, 08193 Bellaterra, Spain}

\date{November 16, 2009}

\begin{abstract}
  Recent works have shown that the domain walls of room-temperature
  multiferroic BiFeO$_3$ (BFO) thin films can display distinct and
  promising functionalities. It is thus important to understand the
  mechanisms underlying domain formation in these films.
  High-resolution x-ray diffraction and piezo-force microscopy,
  combined with first-principles simulations, have allowed us to
  characterize both the atomic and domain structure of BFO films grown
  under compressive strain on (001)-SrTiO$_3$, as a function of
  thickness. We derive a twining model that describes the experimental
  observations and explains why the 71$^{\circ}$ domain walls are
  the ones commonly observed in these films. This understanding
  provides us with a new degree of freedom to control the structure
  and, thus, the properties of BiFeO$_3$ thin films.
\end{abstract}

\maketitle

Magnetoelectric multiferroics exhibit coupled electric and magnetic orders,
which might lead to a variety of novel devices that would benefit from the
fact that the magnetization (polarization) of these materials can be controlled
by means of an electric (magnetic) field~\cite{Ram07}. For practical devices,
multiferroics are preferred in thin film form. Moreover, the strain induced by
the mismatch between the film and the substrate lattice parameters can
sometimes be used to tune the film properties with respect to the
bulk~\cite{Wan03}.

Bismuth Ferrite, BiFeO$_3$ (BFO), is one of the few multiferroics that orders
above 300~K and, thus, one of the most promising ones~\cite{Cat09}. The
ferroelectric properties of BFO are very robust, and it displays record
polarization values of about 100~$\mu$C/cm$^2$. Since the ground state of bulk
BFO is rhombohedral (space group {\sl R3c}), symmetry arguments suggest that
the thin films grown on cubic substrates under compressive epitaxial strain
should be monoclinic (space group {\sl Cm} or {\sl Cc}, depending on whether
the O$_6$ octahedra rotations are clamped by the substrate or not,
respectively). Indeed, several authors~\cite{Xu05,Bea07,Kim08,Jan08} have
reported a monoclinic unit cell that is similar to that of strong
piezoelectric PbZr$_{1-x}$Ti$_{x}$O$_3$ (PZT) with
$x\approx$~0.5~\cite{Noh99}.  The proposed link between the strong
piezoelectricity and the symmetry of the unit cell~\cite{Bel00}, which allows the polarization to rotate, adds to the
interest of BFO films~\cite{Jan08}.

Beyond their intrinsic properties, BFO films are currently receiving renewed
attention because of the novel functionalities observed to occur at domain
walls (DWs). Indeed, recent works have shown that some BFO DWs are highly
conductive~\cite{Sei09}, and that the DW density controls the magnitude of the
(exchange bias) coupling between BFO and other (metallic) layers in complex
heterostructures~\cite{Mar08}. It is thus of prime importance to achieve
control of the domain structures and understand their formation. In contrast,
it is striking to note the scarcity, and lack of agreement, of experimental
information on the atomic structure of the films and its evolution with
thickness~\cite{Xu05,Kim08,Jan09}. Indeed, we believe that a complete picture
of the structure of these films does not exist yet.

We have grown BFO thin films on SrRuO$_3$-buffered SrTiO$_3$ (STO) substrates, and
followed the unit cell distortion as a function of thickness during the first
stages of strain relaxation. Our c/a ratios are consistent with those in
Ref.\onlinecite{Kim08}. Additionally, we have been able to resolve the monoclinic distortion and measure the evolution of the
full unit cell. The comparison of the experimental results with several
structural models simulated {\sl ab initio} allowed us to resolve the
monoclinic space group ({\sl Cc}) and atomic structure, as well as the polarization direction.

Several (001)-oriented BFO thin films with thickness ranging from 12 to 87~nm
were grown on atomically flat, TiO$_2$-terminated (001)-STO substrates with
low miscut angle (0.1$^{\circ}$). Conductive layers of SrRuO$_3$ with a
thicknesses of 5 nm were deposited in between the substrate and the BFO
layer. The BFO films were grown by pulsed laser deposition, assisted by
reflective high energy electron diffraction (RHEED), using a KrF excimer laser
($\lambda$=~248~nm). The deposition was performed at 670$^{\circ}$C in an
oxygen pressure of 0.3~mbar. After deposition, the films were cooled down
slowly to room temperature under an oxygen pressure of 100~mbar.

\begin{figure}
\includegraphics[keepaspectratio=true,width=8 cm,height=6 cm]{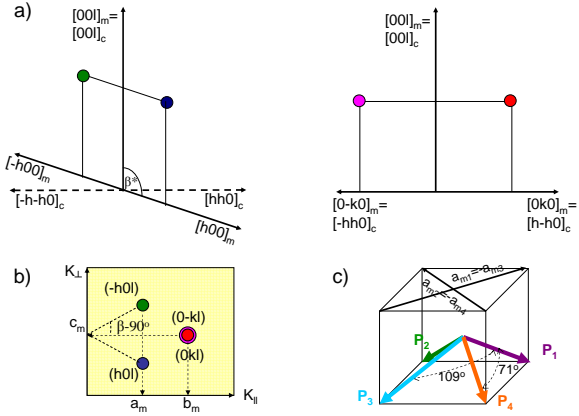}
\caption{a) Monoclinic domains in BFO thin films under compressive strain, in the (h0l)$_m$ (left) and (0kl)$_m$ (right) scattering planes. b) Diffraction map around the (h0l)$_m$ reflections when all four domains are present. c) Directions of the polarization and the monoclinic lattice parameter, a$_m$, for the four down polarized domains}
\end{figure}

The evolution of the crystallographic distortion with thickness was
investigated by mapping the reciprocal space using x-ray diffraction (XRD)
from lab (out-of-plane) and synchrotron (in-plane) sources. Due to the
epitaxy, which fixes the [001] direction in reciprocal space to be
perpendicular to the substrate surface, the twelve possible monoclinic domains
are reduced to four and the reciprocal space maps are significantly simplified
(similar to the case of a crystal under an electric field~\cite{Noh01}). In
particular, if all domains are present, looking at the areas around the
substrate (hhl)$_c$, which corresponds to the (h0l)$_m$ reflection in the
monoclinic structure~\cite{fn1}, one can extract the three lattice parameters
and the monoclinic angle, as sketched in Fig.~1.

\begin{figure*}
\begin{center}
\includegraphics[keepaspectratio=true,width=16 cm,height=8.5 cm]{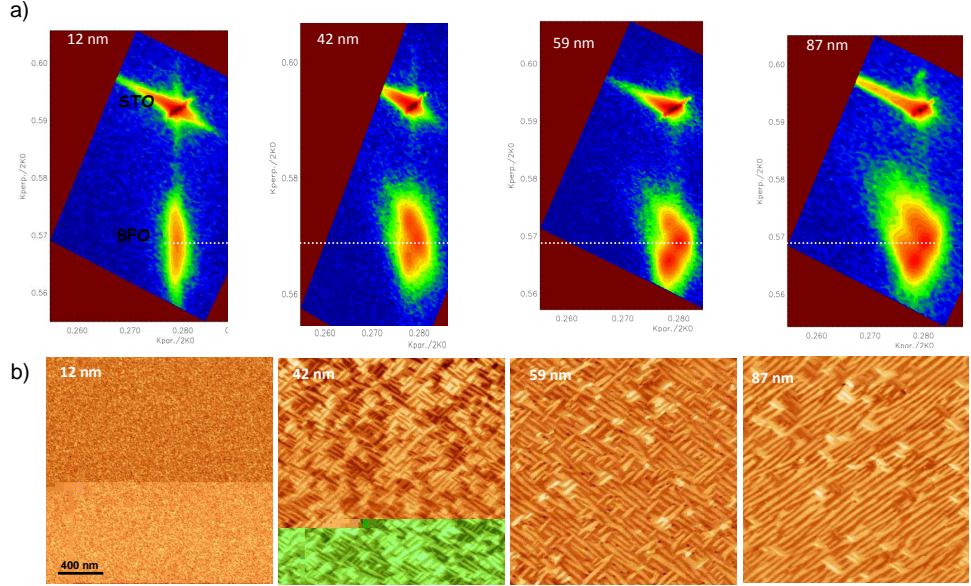}
\caption{a) Reciprocal space maps around the (103)$_m$(= (113)$_{pc}$) reflections for different thicknesses of the BFO films. The axes are in units of 2K$_o$= 4$\pi/\lambda$, with $\lambda$= 1.5405{\AA}. The horizontal line through the maps indicates the out-of-plane reciprocal lattice spacing, which is unchanged in the range of studied thicknesses. b) In-plane piezo-response images of the same films.}
\end{center}
\end{figure*}

Typical reciprocal space maps (RSMs) around the (113)$_c$ STO
substrate reflections for ultrathin ($<$~18~nm) and thin ($>$~18~nm) films
are shown on Fig.~2a. The RSMs of the thinnest BFO films display a
broad (113)$_{c}$ peak (using the pseudo-cubic notation), at the same
K$_{\parallel}$ (in-plane component of the scattering vector) of the substrate, showing that they are fully coherent with the
substrate. The FWHM of these films agrees with what is expected for their thicknesses. There is thus no indication of
unresolved splitting. The RSMs of thicker
films display a splitting of the (113)$_{pc}$ BFO peak, as expected
(see Fig.~1). The monoclinic lattice parameters extracted from these
patterns are plotted in Fig.~3.
Interestingly, ${c_m}$ shows no changes with increasing
thickness. This is in agreement with the report by Kim {\sl et
  al}.~\cite{Kim08}, who showed that the lattice parameters of the
strained films are constant below $\sim$~100~nm, a puzzling and
unexplained result. However, for thickness above 18~nm we observe a
splitting of the in-plane parameter values and a $\beta$$\neq$90$^{\circ}$, characteristic of a
monoclinic distortion. Figure~3 reveals a gradual increase of
the monoclinic distortion ${a_m-b_m}$ with thickness. In addition, grazing incidence XRD (GID) has shown that the
in-plane pseudo-cubic angle, $\gamma_{pc}$ is, indeed, different from
the out-of-plane angle $\beta$, and that such a difference decreases
with increasing thickness ($\gamma_{pc}= \beta$ in the relaxed structure). Interestingly, the deviation of ${a_m}$ and ${b_m}$ from the
value of 2$\times d_{(110)}$ (i.e., the fully coherent case) is symmetric.

First-principles simulations~\cite{fn2} allowed us to ratify these results and gain further insight into the atomic structure of the BFO films. It has been
shown that ferroelectric thin films can be successfully studied by simulating
the corresponding bulk material subject to elastic boundary conditions that
mimic the epitaxial constraints imposed by the substrate. In this work we
extend such an approach to make a distinction between the cases of ultra-thin
and thin films, for which we consider different elastic constraints. More
precisely, in the ultra-thin (uth) case we assumed the film is strongly
clamped by the substrate, and impose $a_{pc}=b_{pc}=a_{\rm STO}$ and
$\alpha_{pc}=\beta_{pc}=\gamma_{pc}$=90$^{\circ}$. In contrast, in the thin
(th) case we only imposed that the in-plane area be constrained to be $a_{\rm
  STO}^2$. This allowed us to model the “ultra-thin” to “thin” transition
evidenced by the experimental results of Fig.~3. On the other hand, in our
simulations we considered two structural models, with and without rotations of
the O$_6$ octahedra, which correspond, respectively, to the {\sl Cc} and {\sl
  Cm} space groups.

Our simulations clearly indicate that the BFO films present significant O$_6$
rotations and thus the {\sl Cc} space group. Indeed, when allowing for O$_6$
rotations we computed $a_{pc}^{\rm th}/a_{pc}^{\rm uth}$=1.0016 and
$b_{pc}^{\rm th}/b_{pc}^{\rm uth}$=0.9981 for the splitting of in-plane
lattice parameters, in good agreement with the experimental values of 1.0015
and 0.9980 derived from the data in Fig.~3. In contrast, when the O$_6$
tiltings are clamped in the simulations, we obtained $a_{pc}^{\rm
  th}/a_{pc}^{\rm uth}$=1.0029 and $b_{pc}^{\rm th}/b_{pc}^{\rm
  uth}$=0.9968. The $c_{pc}/a_{pc}$ ratios follow the same pattern: the value
computed with (without) O$_6$ tiltings is about 1.07 (1.17), to be compared
with the experimental result of approximately 1.04, which strongly suggests
that even in the thinnest films the O$_6$ rotations are not fully
clamped by the substrate. These results provide a justification to first-principles studies of
monoclinic BFO films in which a structural model with O$_6$ rotations is
adopted (see, e.g., Ref.~\onlinecite{hatt09}). Additionally, for the
calculated monoclinic angle we obtained $\beta$=90.4$^{\circ}$, which is
perfectly compatible with our experimental results, and we computed
$c_{pc}^{\rm th}/c_{pc}^{\rm uth}$=1.0005, in agreement with our experimental
observation that the $c_{pc}$ lattice constant is weakly dependent on
thickness. Finally, the computed polarization is very weakly affected by the
uth-to-th transition: We obtained $P$=~90~$\mu$C/cm$^2$, with in-plane and
out-of-plane components of 59~$\mu$C/cm$^2$ and 68~$\mu$C/cm$^2$,
respectively. The polarization forms an angle of about 11.6$^{\circ}$ with the
body diagonal of the pseudocubic cell, being rotated towards the [001]
direction.

\begin{figure}
\includegraphics[keepaspectratio=true,width=7 cm,height=14 cm]{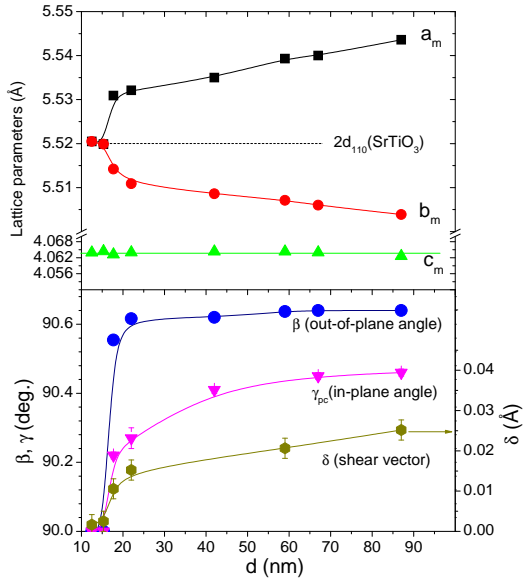}\\
\caption{Evolution of monoclinic lattice parameters and shear displacement as a function of thickness.}
\end{figure}

Let us now describe the evolution of the domain structure. Analysis of the RSMs in Fig.~2 shows that the domain
walls that prevail are the 71$^{\circ}$ ones (see Fig.~1c) in agreement with
previous reports~\cite{Chu07}. This is confirmed by grazing incidence
diffraction (not shown here), as well as from PFM images (Fig.~2b). Out-of-plane PFM measurements show that all the films are polarized down, also in
agreement with previous reports~\cite{Chu07}. In Fig.~2b, in-plane PFM
(IP-PFM) images of the same films are shown. In agreement with the XRD data, we
observe a clear evolution of the domain pattern. For the thinnest films, no
contrast is detected on the IP-PFM images. IP-PFM images for intermediate films show a clear stripe-like pattern. These stripes indicate four
polarization variants, which are in good agreement with rhombohedral-like
monoclinic distortions~\cite{Zav06}(see figure 1c) and with the RSM maps. We observe that the
number of variants decreases from four to two variants with further increasing
thickness, allowing for longer stripes for the thicker films. Several works
have already shown two-variant stripe domains for (001)-oriented BFO films, by
using miscut STO substrates~\cite{Das06,Jan09} or orthorhombic
substrates~\cite{Chu06}. The origin of this reduction of polarization
orientations in BFO films was reported to be the step-flow growth and the
substrate anisotropy, respectively. Since the growth mode as well as the
substrate miscut in all our films are the same, our results point to yet a
different mechanism.

All this evidence fits a simple but powerful model by which the domain
formation enables and controls the monoclinic distortion of the unit cell. Figure~4a shows how twinning reduces the in-plane
strain introduced by the pseudo-cubic angle, $\gamma_{pc}$ (characteristic of
the monoclinic distortion). Two pairs of twins, coherent along [100] (v$_1$) or along [010] (v$_2$), can form. It can also be seen that, in order to do that, the
in-plane lattice parameters of the film, $a_f$ and $b_f$, deviate equally from
the fully strained values of $a_{srt} = b_{str} = \sqrt{2}a_{\rm STO}$, i.e.,
$\vec{a}_f=\vec{a}_{str}-\vec{\delta}$ and
$\vec{b}_f=\vec{b}_{str}-\vec{\delta}$, as sketched in Fig.~4b. The magnitude
of the shear vector $\vec{\delta}$, therefore, determines both $a_f$ and
$b_f$, which split {\sl symmetrically} with increasing thickness (see
Fig.~3). As a result, the in-plane area of the film is unchanged with respect
to the fully coherent film, which in turn seems compatible with the
observation that the $c_m$ lattice parameter does not vary during strain
relaxation (for thicknesses up to 100~nm). A more subtle result of this
relaxation process is that the symmetry of film unit cell is actually lower
than monoclinic; indeed, it can be seen from Fig.~4b that the angle between
$\vec{a}_f$ and $\vec{b}_f$ is given by $\gamma_f =
\cos^{-1}(\delta^{2}/a_{f}b_{f})$, and thus different from 90$^{\circ}$. A very similar twining mechanism with symmetry lowering has been found in thin films of
TbMnO$_3$ grown on (001)-STO substrates~\cite{Sri09}, which suggests it may be
typical of low symmetry perovskites on cubic substrates.
As observed in figure 4a), the two pairs of variants, 90$^{\circ}$ rotated from each other, are in agreement with the PFM maps of the 42 and 59 nm films and give rise to both 71$^{\circ}$ and 109$^{\circ}$ walls. Even though both nucleate with equal probability in the growing film, on a low miscut substrate, because of the relatively large strain energy store at the boundary between them, for thicker films (and therefore larger strain energy at those boundaries) one of the two variants will be preferred, as observed in the thicker 89~nm film. In the presence of a substrate miscut, the steps can indeed determine which of the two variant is present~\cite{Jan09}, but this will happen in exact substrates provided that the films have enough time to relax.

In summary, we have observed clear trends in the evolution with thickness of
the structure and microstructure of BiFeO$_3$ films on (001)-SrTiO$_3$
substrates. We have shown that the lattice parameters and the film symmetry do
not result simply from the mismatch with the substrate, but also from the
occurrence of a particular twinning that allows for the observed monoclinic
distortion. This twinning model explains why the 71$^{\circ}$ domain walls are so often
observed in atomically flat films on (001)-SrTiO$_3$. Such an effect
provides us with a new degree of freedom for tuning the structural and
physical properties of the thin films. Our results suggest that the physics
behind the effects of epitaxial strain is richer than usually thought, and
that traditional thermodynamic phase diagrams and first-principles models need
to be complemented with knowledge of the domain structure in order to reach a
full understanding of the materials behavior.

\begin{figure}
\includegraphics[scale=0.35]{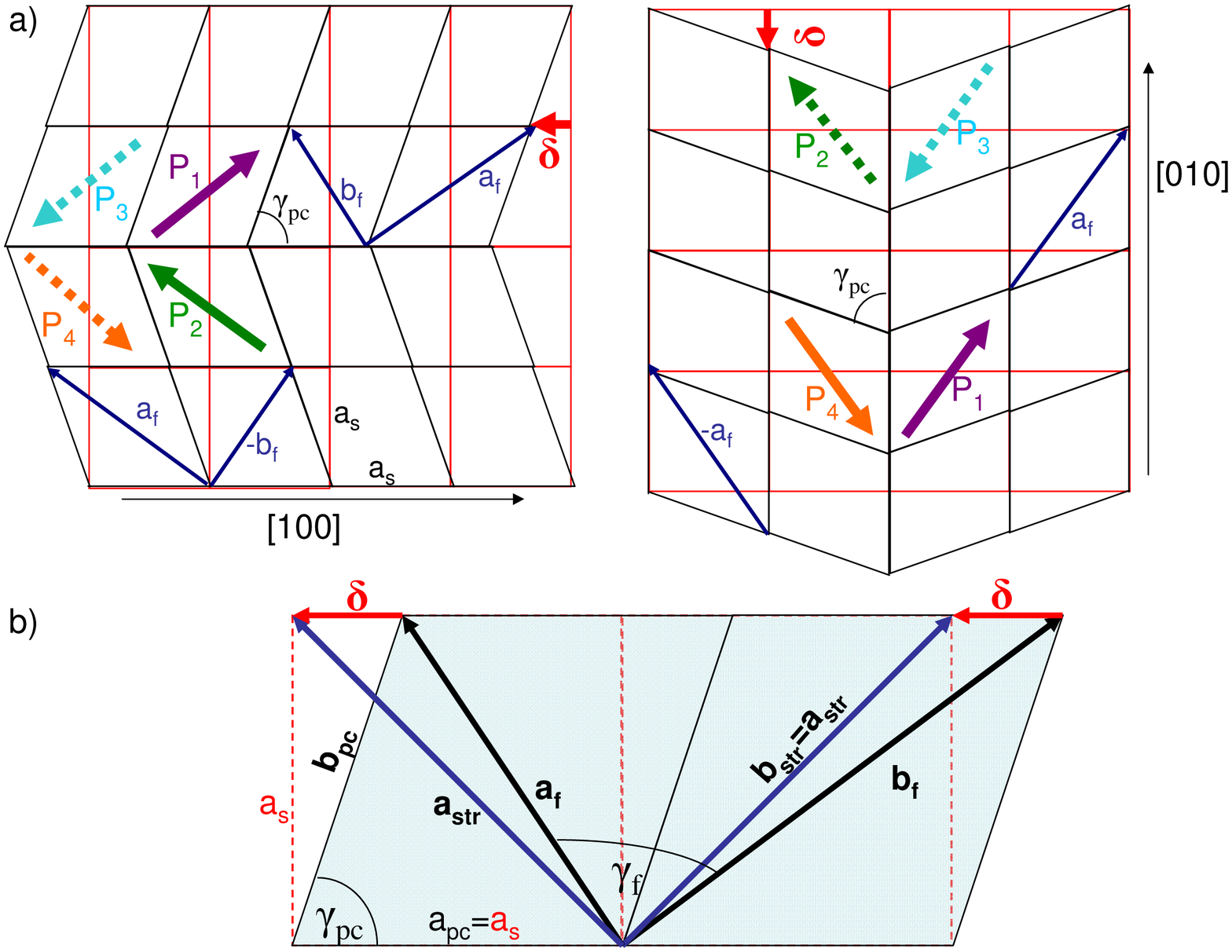}\\
\caption{a) The two types of twins present in the films, each including 71$^0$ walls: v$_1$ (left) is coherent along [100] and v$_2$ (right) is coherent along [010]. b) Detail of the film distortion.}
\end{figure}

We are grateful to Gijsbert Rispens and Sriram Venkatesan for useful discussions. This is work was partly funded by MaCoMuFi (Grant No. STREP\_FP6-03321), the Spanish Government (Grants No. FIS2006-12117-C04-01 and
No. CSD2007-00041) and the Dutch agencies NWO and FOM. We made use of the facilities of the BSC-CNS and CESGA
supercomputing centers.

\

\end{document}